\documentclass[fleqn,twoside]{article}
\usepackage{epsfig}
\usepackage{amsmath}
\usepackage{espcrc2}
\usepackage{float}

\usepackage{graphicx}
\usepackage[figuresright]{rotating}
\def\td#1{\widetilde{\delta}\left(#1\right)}
\def\thd#1{\tilde{\theta}\left(#1\right)}

\def\nn{\nonumber}
\def\Eq#1{Eq.~(\ref{#1})}

\title{Towards a Loop--Tree Duality at Two Loops and Beyond }

\author{I. Bierenbaum\address[IFIC]{Instituto de F\'{\i}sica Corpuscular,
    \\ Universitat de Val\`{e}ncia -- Consejo Superior de Investigaciones
    Cient\'{\i}ficas, \\ Apartado de Correos 22085, E-46071 Valencia,
    Spain}\thanks{This work was supported by the Ministerio de Ciencia e
    Innovaci\'on under Grant No. FPA2007-60323, by CPAN (Grant
    No. CSD2007-00042), by the Generalitat Valenciana under Grant
    No. PROMETEO/2008/069, by the European Commission under contract FLAVIAnet
    (MRTN-CT-2006-035482), and by INFN-MICINN agreement under Grants
    No. ACI2009-1061 and FPA2008-03685-E.}}
       
\begin{document}

\begin{abstract}
We present an extension of the duality theorem, previously defined by
S.~Catani {\it et al} on the one--loop level, to higher loop orders. The
duality theorem provides a relation between loop integrals and tree--level
phase--space integrals. Here, the one--loop relation is rederived in a way
which is more suitable for its extension to higher loop orders. This is shown
in detail by considering the two--loop N--leg master diagram and by a short
discussion of the four master diagrams at three loops, in this sketching the
general structure of the duality theorem at even higher loop orders.
\vspace{1pc}
\end{abstract}

\maketitle

\section{INTRODUCTION}

Theoretical predictions of high precision for background and signal
multi--particle hard scattering processes in the Standard Model (SM) and
beyond, are mandatory in the era after the start of the LHC. To achieve this
high level of precision, one needs to compute cross sections at
next-to-leading order (NLO), or even next-to-next-to leading order (NNLO).
However, going to higher orders in perturbation theory, as well as increasing
the number of external particles for the consideration of multi--particle
processes, both lead to an increase in complexity of the calculations and of
the challenges one has to face in doing the numerical evaluations.  In recent
years, important efforts have been devoted to developing efficient methods for
the calculation of multi--leg and multi--loop diagrams, realized, e.g., by
unitarity--based methods or by traditional Feynman diagram approaches,
\cite{Binoth:2010ra}.

The computation of cross sections at NLO (or NNLO) requires the separate
evaluation of real and virtual radiative corrections, which are given in the
former case by multi--leg tree--level and in the latter by multi--leg loop
matrix elements to be integrated over the multi--particle phase--space of the
physical process.  The loop--tree duality at one--loop presented in
Ref.~\cite{Catani:2008xa}, as well as other methods relating one--loop and
phase--space integrals, \cite{Soperetal,Kilian:2009wy,Moretti:2008jj}, recast
the virtual radiative corrections in a form that closely parallels the
contribution of the real radiative corrections. The use of this close
correspondence is meant to simplify calculations through a direct combination
of real and virtual contributions to NLO cross sections.  Furthermore, the
duality relation has analogies with the Feynman Tree Theorem (FTT),
\cite{Feynman:1963ax,F2}, but offers the advantage of involving only single
cuts of the one--loop Feynman diagrams.

In this talk, the extension of the loop--tree duality theorem derived in
Ref.~\cite{Catani:2008xa} to higher loop orders is described, with the aim of
extending the duality method to the computation of cross sections at NNLO or
even higher orders. This has been explained in detail in
Ref.~\cite{HIGHERLOOPS}. The higher--order dual representations described here
are valid as long as only single poles are present when the residue theorem is
applied (cf.\linebreak Section~3). At two-- or higher loop orders, however,
higher order poles might appear which require a separate treatment which is
beyond the scope of the current talk.

\section{THE DUAL PROPAGATOR $G_D$}

We first state some necessary definitions and introduce the basic formulae
used in the following, leading to the main relation \Eq{eq:GAinGDGeneralN}
which is fundamental for the extension of the duality theorem to higher loop
orders.

The integrals considered are given in $d$
dimensions, where the following short--hand notation is used:
\begin{equation}
\int_{\ell_i} \, \cdots  \equiv
- i \, \int \frac{d^d \ell_i}{(2\pi)^d} \, \cdots~.
\end{equation}
The FTT as well as the duality theorem rely on the pole
structure of the propagators of a given Feynman diagram. For the
Feynman propagator and the advanced propagator defined by:
\begin{eqnarray} 
G_F(q_i) &=& \frac{1}{q^2-m^2+i0}\;, \\
G_A(q_i) &=& \frac{1}{q^2-m^2-i0\,q_{i,0}}~,
\end{eqnarray}
with $q$ being the $d$--dimensional four momentum, whose energy (time
component) is $q_0$, the poles in the complex $q_0$--plane are placed at:
\begin{eqnarray}
\label{fpole}
q^F_{i,0} &=& \pm  {\sqrt {{\bf q}_i^2 -m_i^2-i0}}~~,
\\
q^A_{i,0} &\simeq& \pm  {\sqrt {{\bf q}_i^2 -m_i^2}} +i0 \;\;~.
\end{eqnarray}
The definition of the propagators includes the appearance of masses which have
no effect on the derivation of the duality relation as long as these masses
are real. The question of the occurrence of real and complex masses in the
propagators and their effect on the derivation of the duality theorem, has
been discussed in \cite{Catani:2008xa}. Hence, for the advanced propagator,
both poles have a positive imaginary part, and thus both lie above the real
axis. The poles of the Feynman propagator on the other hand lie above and
below the real axis, depending on the energy being positive or negative. In
addition to these propagators, we will encounter a so--called dual propagator
of $q_j$ with respect to $q_i$ which is defined as, \cite{Catani:2008xa}:
\begin{equation}
G_D(q_i;q_j) = \frac{1}{q_j^2 -m_j^2 - i0 \,\eta (q_j-q_i)}~.
\end{equation}
The auxiliary vector $\eta$ is a a {\em future--like} vector,
\begin{equation}
\label{etadef}
\eta_\mu = (\eta_0, {\bf \eta}) \;\;, \;\; \quad \eta_0 \geq 0, 
\; \eta^2 = \eta_\mu \eta^\mu \geq 0 \;\;,
\end{equation}
i.e.,~a $d$--dimensional vector that can be either light--like $(\eta^2=0)$ or
time--like $(\eta^2 > 0)$ with positive definite energy $\eta_0$.  
Note that if both momenta $q_i$ and $q_j$ depend on the same integration
momentum, the $i0$--prescription of this propagator only depends on external
momenta and hence is integration--momentum free. This is an important
property, since in this case the pole of the considered propagators is at a
well--defined location, either above or below the real axis, depending on the
overall sign of the $i0$--part. In particular, one avoids the
occurrence of branch cuts, which would be introduced in the case of an
integration--momentum dependent $i0$--prescription.

Using the principal value identity, one can infer relations between the
various propagators:
\begin{eqnarray}
G_{A}(q_i) = G_{F}(q_i)+\td{q_i} \;, 
\end{eqnarray}
and
\begin{eqnarray}
&&\hspace{-0.65cm}\td{q_i} G_D(q_i;q_j) \nn \\
&&\hspace{-0.65cm}=
\td{q_i} \Bigl[ G_F(q_j) + \thd{q_j-q_i} \; \td{q_j} \Bigr]~,
\label{dovsfp}
\end{eqnarray}
with
\begin{eqnarray}
\thd{q}\;&:=&\theta(\eta q), \\
\td{q_i}&:=&2 \pi \, i \,
\theta(q_{i,0}) \, \delta(q_i^2-m_i^2) \nn \\
&=& 2 \pi \, i \, \delta_+(q_i^2-m_i^2)~
\end{eqnarray}
where the index $+$ of the delta function refers to
the on--shell mode with positive definite energy, $q_{i,0}\geq 0$.

\begin{figure*}
\begin{center}
  \epsfig{file=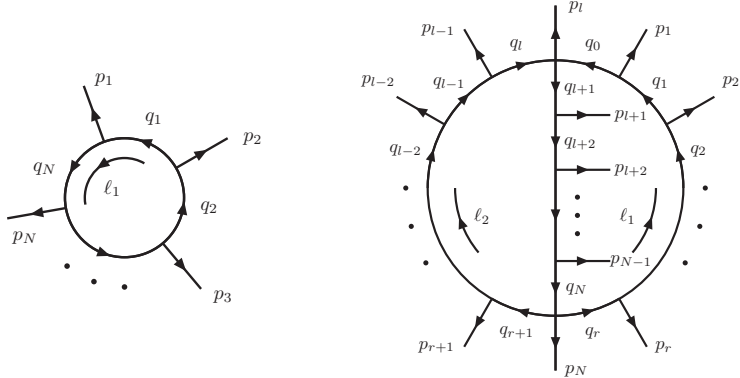,height=5cm}
\end{center}
\vspace*{-6mm}
\caption{\label{f2loop} {\em Momentum configuration of the one--loop and
    two--loop $N$--point scalar integral. Note, that in the two--loop case
    $p_l$ and $p_N$ can be equal to zero and hence the number of internal
    lines can differ from the number of external momenta.}}
\end{figure*}

In the following, it will be important not only to consider single
propagators, but complete sets of them. We therefore define propagator
functions of a set $\alpha_k$ of internal momenta, with $q_i,q_j\in\alpha_k$:
\begin{eqnarray}
\label{eq:multi}
G_{F(A)}(\alpha_k) &:=& \prod_{i \in \alpha_k} G_{F(A)}( q_i)~,\\
G_D(\pm \alpha_k)\; &:=& \sum_{i \in \alpha_k} \, \td{\pm q_{i}} \, 
\prod_{\substack{j \in \alpha_k \\ j \neq i }} \, G_D(\pm q_i;\pm q_j)~. \nn \\
\label{defdual}
\end{eqnarray}
The minus sign in the above definition has the following meaning: Given a set
of momenta $q_i\in\alpha_k$, $-\alpha_k$ denotes the same set of momenta,
where the sense of momentum flow is reversed $q_i \rightarrow -q_i$, $\forall
i\in\alpha_k$. For $\alpha_k = \{i\}$ given by a single momentum, we obtain
$G_D(\pm \alpha_k)=\td{\pm q_i}$.

In analogy to the definitions based on single momenta, we also find a
relation between these propagators of sets of momenta $\alpha_k$:
\begin{equation}
G_A(\alpha_k) = G_F(\alpha_k) + G_D(\alpha_k)~.
\label{eq:relevant}
\end{equation}
This is, however, a non--trivial relation, considering \Eq{dovsfp} and has
been proven in the Appendix of Ref.~\cite{HIGHERLOOPS}. In fact, this relation
is the cornerstone for the duality relations between loops and trees derived
in the following.

As a last step in this section, we derive a formula to express $G_D(\alpha_k)$
in terms of chosen subsets of $\alpha_k$.  Consider thus $\alpha_k$ as a
unification of various subsets, $\alpha_k \equiv \beta_1 \cup ... \cup
\beta_N$. Using \Eq{eq:relevant} and \Eq{eq:multi}, we find:
\begin{eqnarray}
\label{eq:GAinGDGeneralN}
&&\hspace{-0.9cm}
G_D(\beta_1 \cup \beta_2 \cup ... \cup \beta_N) \nn \\
&=& \hspace{-0.2cm}
\sum_{\substack{\beta_N^{(1)} \cup     \beta_N^{(2)}\\  = \alpha_k}} \,
\prod_{\substack{i_1\in \alpha_k^{(1)}}} \, G_D(\alpha_{i_1}) \,
\prod_{\substack{i_2\in \alpha_k^{(2)}}} \,G_F(\alpha_{i_2})\,.  \nn \\
\end{eqnarray}
The sum runs over all partitions of $\alpha_k$ into exactly two blocks
$\alpha_k^{(1)}$ and $\alpha_k^{(2)}$ with elements $\beta_i,\linebreak i\in
\{1,...,N\}$, where, in contrary to the usual definition, we include the case:
$\beta_N^{(1)} \equiv \beta_N$, $\beta_N^{(2)} \equiv \emptyset$.  Using this
equation for different choices of subsets will be the major step to
demonstrate the majority of the following relations.

\section{DUALITY AT ONE--LOOP}

We summarize first the duality relation as derived in
Ref.~\cite{Catani:2008xa}, before reformulating it in a way, which is more
suitable for its extension to higher loop orders. Consider a one--loop scalar
Feynman diagram with $N$ external legs as shown on the left--hand side of
Fig.~\ref{f2loop}.
All external momenta are taken as outgoing and defined modulo $N$. The
integration momentum is $\ell_1$, and the internal momenta $q_i$ are defined
as $q_i = \ell_1 + p_{1,i}~,$
where $p_{i,j}=p_i+p_{i+1}+\ldots+p_j$.  The diagram is represented by the
following function:
\begin{equation}
\label{Ln}
L^{(1)}(p_1, p_2, \dots, p_N) =
\int_{\ell_1} \, \prod_{i=1}^{N} \, G_F(q_i)~.
\end{equation}
Closing the integration contour at infinity in the direction of the negative
imaginary axis, according to the Cauchy theorem, one picks up one pole from
each of the $N$ Feynman propagators. In Ref.~\cite{Catani:2008xa}, it was
shown that the residue of these poles is given by:
\begin{equation}
\label{resGi}
{\rm Res}  [   G_F(q_i) ]_{{\rm Im}(q_{i,0}) < 0} 
= \int d \ell_{1,0} \, \delta_+(q_i^2-m_i^2)~.
\end{equation}
Since we are taking residues at poles in the complex plane, this shifting to
the location of residues for one propagator modifies the imaginary part of the
remaining propagators in the original integral from Feynman propagators to
dual propagators and we thus obtain:
\begin{eqnarray}
\label{oneloopduality}
&&
\hspace{-0.95cm}L^{(1)}(p_1, p_2, \dots, p_N) \nn \\
\hspace{-0.95cm}
&=&
 - \sum \, \int_{\ell_1} \; \td{q_i} \,
\prod_{\substack{j=1 \\ j\neq i}}^{N} \,G_D(q_i;q_j)~.
\end{eqnarray}
All propagators and hence all momenta $q_i,q_j$ depend on only one integration
momentum, $\ell_1$, and therefore this dependence drops out in the difference
of the two, causing the $i0$--prescription in \Eq{oneloopduality} to solely
depend on external momenta.

We will now use a different approach to derive this formula and go back to
\Eq{eq:relevant}: Let us assume that the set $\alpha_k$ contains all internal
momenta $q_i$ of the one--loop integral. We further take the integral on both
sides of \Eq{eq:relevant} and close the contour as described before in the
direction of the negative imaginary axis up to infinity: 
\begin{equation}
\int_{\ell_1}  G_A(\alpha_k) = 
\int_{\ell_1}  G_F(\alpha_k) + \int_{\ell_1} G_D(\alpha_k)~.
\end{equation}
The first integral on the right hand side is the original Feynman integral at
one--loop, whereas the integral over advanced propagators has no poles in this
integration area, and thus vanishes, leaving us with the identity:
\begin{equation}
\label{eq:simpledual}
L^{(1)}(p_1, p_2, \dots, p_N)
= - \int_{\ell_1} G_D(\alpha_1)~.
\end{equation}
Comparing \Eq{oneloopduality} to \Eq{eq:simpledual}, we notice that, via
\Eq{defdual}, they are obviously the same. \Eq{eq:simpledual} is true for any
set of momenta which depends on the same loop momentum with respect to which
the integral is taken, and is the application of the duality relation to the
set $\alpha_1$.

Note that if applying \Eq{eq:GAinGDGeneralN}, with all subsets given by
the single inner momenta of the one--loop integral, $\alpha_i=\{i\}$, one
immediately rederives the FTT at one--loop.

\section{DUALITY AT TWO LOOPS}

The main goal at two-- and higher loop order is to find a formula similar to
the one in the one--loop case, with on the one hand the number of cuts
preferably equal to the number of loops and furthermore an $i0$--prescription
of the dual propagators, which depends on external momenta only. At two loops,
the general $N$--leg master diagram is shown on the right--hand side of
Fig.~\ref{f2loop}.  Again, all momenta are taken as outgoing, while we now
have two integration momenta $\ell_1$ and $\ell_2$. Three so--called ``loop
lines'' $\alpha_k$ are denoted according to the set of internal momenta which
they are labeling, as indicated in Fig.~\ref{f2loop}:
\begin{eqnarray}
\label{lines}
&&\hspace{-0.65cm}\alpha_1= \{0,1,...,r\}~, \nn \\
&&\hspace{-0.65cm}\alpha_2= \{r+1,r+2,...,l\}~, \nn \\
&&\hspace{-0.65cm}\alpha_3= \{l+1,l+2,...,N\}~. 
\end{eqnarray}

We can now directly make use of the results obtained in the one--loop case in
the following way: We start by applying the duality to the first loop momentum
$\ell_1$ and therefore, using \Eq{eq:simpledual}, to the related sets $\alpha_1
\cup \alpha_3$:
\begin{eqnarray}
&&\hspace{-0.7cm}L^{(2)}(p_1, p_2, \dots, p_N) \nn \\
&&\hspace{-0cm}= - \int_{\ell_1} \, \int_{\ell_2} \,
G_D(\alpha_1 \cup \alpha_3) \, G_F(\alpha_2)~. 
\end{eqnarray}
No matter how we name the loop lines, one of them will always depend on both
loop momenta, $\alpha_3$ in this case, and thus $G_D(\alpha_1 \cup \alpha_3)$
is an expression that contains some terms, which are not free of integration
momenta in their $i0$--prescription. However, this can be restored by using
\Eq{eq:GAinGDGeneralN} which allows us to express this dual in terms of its
subsets, for which we choose the lines $\alpha_1$ and $\alpha_3$. A dual
function of only one of these lines as defined in \Eq{lines} is obviously
integration--momentum independent in the desired manner. Therefore, we will
always try to represent our results in terms of $G_D(\alpha_k)$, $k=1,2,3$ as
defined in \Eq{lines}. These are the maximal sets of propagators with
momentum--independent $i0$--prescription available.

For the case of the two--loop master integral considered here, we obtain:
\begin{eqnarray}
\label{eq:L2firstloop}
&&\hspace{-0.7cm}L^{(2)}(p_1, p_2, \dots, p_N) = \nn \\
&&\hspace{-0.7cm}- \int_{\ell_1} \, \int_{\ell_2} \,
\left\{ G_D(\alpha_1) \, G_D(\alpha_3)
+ G_D(\alpha_1) \, G_F(\alpha_3) \right.
\nn \\[0.2em]
&&\hspace{1cm}
\left.
 + G_F(\alpha_1) \, G_D(\alpha_3) \right\} \, G_F(\alpha_2)~. 
\end{eqnarray}
The first term of the integrand on the right--hand side of \Eq{eq:L2firstloop}
is the product of two dual functions, and therefore already contains double
cuts.  We do not modify this term further. The second and third terms of
\Eq{eq:L2firstloop} contain only single cuts and we thus apply the duality
theorem again, i.e., use \Eq{eq:simpledual} for $\ell_2$. A subtlety arises at
this point since due to our choice of momentum flow, $\alpha_1$ and $\alpha_2$
appearing in the third term of \Eq{eq:L2firstloop}, flow in the opposite
sense. Hence, in order to apply the duality theorem to the second loop, we have
to reverse the momentum flow of one of these two loop lines. We choose to
change the direction of $\alpha_1$, namely $q_i\to -q_i$ for $i \in
\alpha_1$. Thus, applying \Eq{eq:simpledual} to the last two terms of
\Eq{eq:L2firstloop} and expanding all parts in terms of the single loop lines
of \Eq{lines} leads to
\begin{eqnarray}
\label{AdvDualstar}
&&\hspace{-0.7cm}
L^{(2)}(p_1, p_2, \dots, p_N)  \nn \\[0.2em]
&&\hspace{-0.7cm} = \int_{\ell_1} \int_{\ell_2} \, \left\{
  G_D(\alpha_1)  \, G_D(\alpha_2) \, G_F(\alpha_3)\right. \nn \\[0.2em]
&& \hspace{0.5cm} \left. 
 + G_D(-\alpha_1) \, G_F(\alpha_2) \, G_D(\alpha_3)\right. \nn \\[0.2em] 
&& \hspace{0.5cm} \left. 
+ G^*(\alpha_1)  \, G_D(\alpha_2) \, G_D(\alpha_3) \right\}~, 
\end{eqnarray}
where
\begin{equation}
\label{Gstar1}
G^*(\alpha_k) \equiv G_F(\alpha_k) + G_D(\alpha_k) + G_D(-\alpha_k)~.
\end{equation}
This is the main result for the two--loop diagram. 

In \Eq{AdvDualstar}, the $i0$--prescription of all dual propagators depends on
external momenta only. Through \Eq{Gstar1}, however, \Eq{AdvDualstar} contains
also triple cuts, given by the contributions with three $G_D(\alpha_k)$. The
triple cuts are such that they split the two--loop diagram into two
disconnected tree--level diagrams, however, there is no more than one cut per
loop line $\alpha_k$. At one loop, there is a single loop line $\alpha_1$, and
we cannot introduce more than one single cut in the dual representation of a
diagram. For a higher number of loops, we expect to find at least the same
number of cuts as the number of loops, and topology dependent disconnected
tree diagrams built by cutting up to all the loop lines $\alpha_k$. This is a
natural consequence of the application of \Eq{eq:GAinGDGeneralN}. We
investigate this question again at three loops in the next section.

Note that using \Eq{eq:relevant}, $G^*(\alpha_k)$ can also be expressed as
\begin{equation}
\label{Gstar2}
G^*(\alpha_k) = G_A(\alpha_k) + G_R(\alpha_k) - G_F(\alpha_k)~,
\end{equation}
which contains no cuts, although the imaginary prescription of
the advanced and retarded propagators still depends on the 
integration loop momenta. 

As in the one--loop case, by using \Eq{eq:GAinGDGeneralN} with subsets
$\alpha_k=\{i\}$ given by all inner momenta, we can also in this case infer
a FTT at two loops, ranging from double--cuts to $N$--tuple cuts, with $N$ the
total number of internal lines, propagators respectively,
cf. \cite{HIGHERLOOPS}.

\section{BEYOND TWO LOOPS}

\begin{figure*}[t]
\begin{center}
\epsfig{file=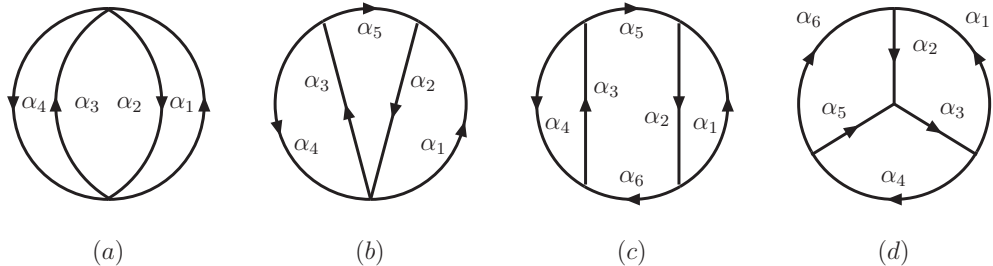,height=3.5cm}
\end{center}
\vspace*{-7mm}
\caption{\label{f3loop} {\em Master topologies of three--loop scalar
    integrals. Each internal line $\alpha_i$ can be dressed with an arbitrary
    number of external lines, which are not shown here.  }}
\end{figure*}

From the derivation of the formulae in the last section, it is clear that one
can go to higher loop orders by iteratively applying the duality
\Eq{eq:simpledual} to the occurring loops. In Ref.~\cite{HIGHERLOOPS}, all
four scalar master diagrams at three loops are considered as given in
Fig.~\ref{f3loop} and formulae provided for their results. These results are
not unique in the sense that there is some freedom in the choice of
propagators for which the directions of momentum flow are changed in doing the
various cuts. For diagrams (a) to (c) a result can be derived in a very
similar and fast way, while diagram (d), being the only non--planar diagram,
is slightly more involved. As expected, it could be confirmed that the final
result always consists of cuts, with a multiplicity ranging from the number of
loops of the diagram, up to a cut with the multiplicity of the number of loop
lines, where each loop line is cut exactly once, which has the effect of
generating disconnected graphs.

The extension to higher loop orders above three loops seems straightforward,
though more complex topologies might complicate the way of constructing the
final results and its overall structure. However, due to the use of
\Eq{eq:GAinGDGeneralN}, this result should again involve a sum over cuts with
multiplicity starting from the number of loops up to the number of loop lines.

\section{CONCLUSION AND OUTLOOK}

The duality--theorem derived in Ref.~\cite{Catani:2008xa} has been
reformulated in a way, which allows to extend it to higher loop orders. We
explained in detail the steps to obtain a duality relation at two loops,
considering the two--loop scalar master diagram and shortly discussed its
application to the four three--loop master diagrams. Following these examples,
this method provides a straightforward prescription to apply it to even higher
loop orders.

\end{document}